\documentclass[lineno]{jfm}
\RequirePackage{snapshot}
\usepackage[utf8]{inputenc}
\usepackage{amsmath}
\usepackage{amssymb}
\usepackage{booktabs}
\usepackage{color}
\usepackage{graphicx}
\usepackage{epstopdf}
\usepackage{hyperref}
\usepackage{listings}
\usepackage{mathtools}
\usepackage{multicol}
\usepackage{multimedia}
\usepackage{natbib}
\usepackage{xcolor}
\usepackage[inkscapeformat=pdf]{svg}

\newcommand{\dzy}{\ensuremath{\partial \overline{\xi}_{y,y}}}
\newcommand{\calL}{\ensuremath{{\cal L}}}
\newcommand{\calU}{\ensuremath{{\cal U}}}

\newcommand{\Fh}{\ensuremath{\mathit{Fr_h}}}

\newcommand{\Ft}{\ensuremath{\mathit{Fr_t}}}

\newcommand{\Gn}{\ensuremath{\mathrm{Gn}}}
\newcommand{\Lh}{\ensuremath{\mathit{L_h}}}

\newcommand{\Pe}{\ensuremath{\mathit{Pe}}}

\newcommand{\Rb}{\ensuremath{\mathit{Re_b}}}

\newcommand{\Rh}{\ensuremath{\mathit{Re_h}}}

\newcommand{\Rt}{\ensuremath{\mathit{Re_t}}}

\newcommand{\Rey}{\ensuremath{\mathit{Re_\lambda}}}

\renewcommand{\vec}[1]{\mathbf{#1}}

\hypersetup{
  colorlinks = true,
  allcolors = blue,
  citecolor = black,
  linkbordercolor = {white},
}

\title[Passive scalar mixing layers]{Evolution of passive scalar mixing layers in stratified and unstratified homogeneous turbulence}

\author
  [Stephen M. de Bruyn Kops, Peter N. Blossey and James J. Riley]
    {Stephen M. de Bruyn Kops$^1$, Peter N. Blossey$^2$ and \\James J. Riley$^3$}

\affiliation{
         $^1$Department of Mechanical and Industrial Engineering,\\ University
             of Massachusetts Amherst, Amherst, Massachusetts, USA\\
        $^2$Department of Atmospheric and Climate Science, \\University
             of Washington, Seattle, Washington, USA \\
         $^3$Department of Mechanical Engineering, \\University
             of Washington, Seattle, Washington, USA \\
         }

\begin{document}

\maketitle

\noindent
{\small \textsuperscript{\textcopyright}\the\year{} by Stephen M.\ de Bruyn Kops, Peter N. Blossey and James J. Riley}

\vspace*{2ex}
\begin{abstract}
High-resolution large-eddy simulations of decaying stratified and unstratified homogeneous turbulence are used to understand the mixing of passive scalars in stably stratified flows.  Two passive scalar mixing layers, one whose mean gradient is in the vertical direction and the other in the transverse direction, are a model for a plume that is very large relative to the length scale of the velocity.  In the transverse direction, the evolution of the passive scalar is broadly similar in the stratified and unstratified cases, although it does spread slightly faster when stratified.  Also, the intensity of the scalar fluctuations is higher in the stratified case, and the turbulent/non-turbulent interface is more intermittent.  With the mean gradient in the vertical direction, however, the stratified case has almost no mixing because the stratification prevents large-scale stirring. Initially, the stratified passive layer grows until its width is proportional to the vertical integral length of the horizontal velocity, which is itself constrained to maintain the vertical Froude number order one.  After this early growth, there is little additional spreading of the passive scalar.  Modelling of the stratified scalar flux in the transverse direction is done effectively with a one-constant model if the mean profile is known, and a two-constant model if the profile shape must be assumed.  In the latter case, the model is good only if the scalar is in quasi-equilibrium with the velocity field such that the length scale of the scalar can be scaled from the kinetic energy.  In this study, the Prandtl number of the active and passive scalars is 0.7.  It is anticipated that the reverse buoyancy flux resulting from higher Prandtl numbers will affect the passive scalar mixing.

\end{abstract}



\section{Introduction}
The dispersion of particles and other contaminants in the atmosphere, oceans, and technological flows can be of great consequence.  For example, it is often important to understand and predict the behaviour of aerosol particles (such as dust, soot, pollen and volcanic ash), cloud droplets and ice crystals in the upper troposphere and stratosphere.  This is also true of the behaviour of pollutants, microplastics, and various small organisms in the oceans and drinking-water reservoirs.  Understanding and predicting the behaviour of these is made more difficult because all of these flows involve stably stratified turbulence (SST), at least some of the time.  The stable stratification inhibits vertical motion, enables the propagation of internal waves, and greatly modifies the turbulence.  Much has been learned about how turbulence with stable density stratification affects the dispersion and mixing of the density field itself (see, e.g., the review of \cite{caulfield2021}), but much less on the dispersion of contaminants.  The dispersion of particles depends very much on how the particles have been introduced, for example as contrails from an airplane \citep[e.g.,][]{voigt17}, or microplastics in a river outflow \citep[e.g.,][]{diben26}.

A major advance in understanding and predicting the dispersion of particles in non stratified flows was made by Taylor (1921), who considered the dispersion of fluid particles in a homogeneous, isotropic, stationary turbulent flow.  He related the turbulent diffusivity $D$ to the particle velocity autocorrelation function and found that it asymptotes as $D = \langle u^2 \rangle T_L$, where $\langle u^2 \rangle$ is the Lagrangian mean square velocity, and $T_L$ is the Lagrangian velocity integral time scale.  \cite{csanady64}  followed Taylor’s approach and attempted to estimate the effect of stable stratification on the particle velocity autocorrelation and hence on  Lagrangian dispersion.  He found that the standard deviation of the vertical fluid particle displacement tends asymptotically to a constant value due to the suppression of the vertical motion by the stable stratification.  This result was in qualitative agreement with some existing field data.  \cite{osborn72} assumed a balance between the production and dissipation terms in the equation for the mean square temperature fluctuations in the ocean, and found that the temperature diffusivity $D_T$ can be approximated as
$$
D_T = \frac{\epsilon_T}{\left({\partial \left<T\right>} \over {\partial z}\right)^2} \, ,
$$
where $\langle T \rangle$ is the mean temperature profile, and $\epsilon_T = \kappa_T \langle \nabla T \cdot \nabla T \rangle$ is the mean square temperature dissipation rate.  Based upon a result by \cite{lindborgbreth08} regarding vertical particle displacement in a stratified flow, \cite{lindborgfed09} determined that the vertical eddy diffusivity for fluid particles $D$ in a stable stratified fluid can be given as:
$$ D = {\epsilon_p \over N^2} \, ,
$$
where $\epsilon_p$ is the dissipation rate of mean square buoyancy fluctuations, and $N$ is the buoyancy frequency.  By considering the case of the vertical diffusion of a horizontal ‘slab’ of a scalar in a forced, statistically steady, stably-stratified turbulent fluid flow, Lindborg and Fedina were then able to demonstrate good quality in this model by comparing the results of the model with those from direct numerical simulations.  \cite{qian22} show that the assumptions of homogeneity and stationarity can be relaxed if density-sorted coordinates are used \citep{winters95,winters96}.

When information about $\epsilon_p$ is not available, the kinetic energy dissipation rate $\epsilon_k$ can be used to estimate the vertical eddy diffusivity of heat:
$$ D = {\Gamma \epsilon_k \over N^2} \, ,
$$
where $\Gamma = \epsilon_p / \epsilon_k$ is a mixing parameter, whose value is uncertain but is hypothesized to be 1/3 for stratospheric turbulence based on arguments about the flux Richardson number \citep{lilly74}.  While vertical mixing is limited by stratification, constraints on horizontal mixing are much weaker. Estimated horizontal diffusivities in stratified turbulence exceed the vertical diffusivities by orders of magnitude, and their ratio may grow over time as the horizontal length and velocity scales grow \citep{pisso09,schumann12}.  These anisotropic diffusivities are consistent with large horizontal-to-vertical aspect ratios of turbulent patches observed in the stably stratified upper tropsphere \citep{podglajen17}.

While some contaminants are fluids and others particles, we assume the particles act as fluid particles and make the continuum approximation to treat an assemblage of particles as a continuum.  Furthermore, we focus on situations in which the contaminants can be treated as passive scalars.  We then choose
a simple configuration for studying inhomogeneous turbulent mixing, a
passive scalar initialised with a step change in the scalar value across the
flow. Research starting in the 1970s suggests that the scalar
statistics for a mixing layer in homogeneous isotropic turbulence (HIT) at high
P\'eclet and Reynolds numbers can be expected to approach self-similarity in time with departure from self-similarity being consistent with finite P\'eclet number $\Pe$, which introduces additional length and time scales into the flow
\citep{libby75,larue81a,larue81b,lumley86,ma86,debk00a}. In short,
the scalar statistics for a mixing layer in HIT can be well modelled in terms of
the integral length and time scales of the turbulence.

\begin{figure}
\centering
    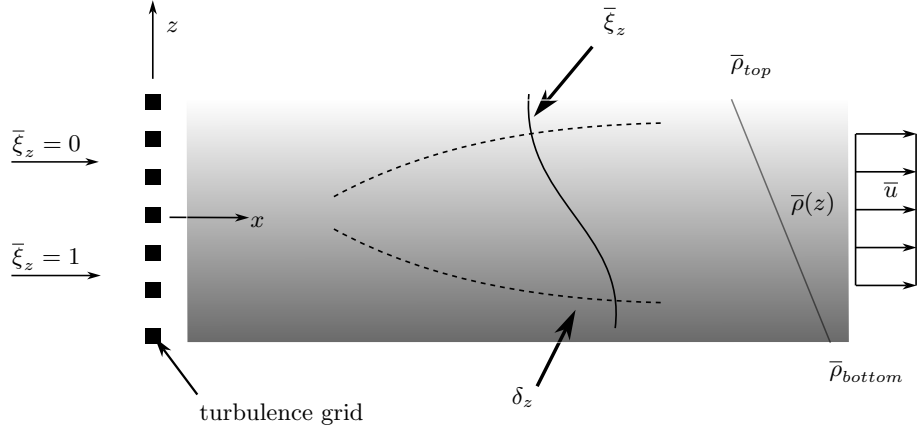
    \caption{Schematic of a passive scalar mixing layer $\overline{\xi}_z$ as if a
      density gradient were suddenly imposed on grid turbulence.  Shading
      depicts the density gradient $d\overline{\rho}/{dz}$.  In the
      simulations, there is a second mixing layer $\overline{\xi}_y(y)$
      oriented out of the page.
    \label{fig:schematic}}
\end{figure}

An open question is what happens when a passive scalar mixing layer (PSML) is
introduced into HIT and subjected to a stabilising density gradient such that the
buoyancy force becomes comparable to the inertial force as the flow decays in
time.  Figure \ref{fig:schematic} is a schematic of this flow configuration.
Whereas in HIT at high $\Pe$ the inertial scales dominate so that
the scalar statistics are nearly self similar, stable stratification
introduces buoyancy scales and also anisotropy, which can be expected to
affect the evolution of the passive scalar.  This is the topic of the research
reported here.

Initially homogeneous and isotropic turbulence that evolves with stable
stratification is one of the simplest configurations of stably stratified
turbulence (SST), which, in turn, informs us about a variety of scenarios in
the ocean, atmosphere, and lakes, as well as in industrial applications.  It
is a model for a flow that starts out sufficiently energetic so that buoyancy
forces are negligible, but then decays so that the vertical Froude number is
order one and the horizontal Froude number is order one or less, indicating the strong effect of density stratification.
If the initial Reynolds number is high then, after about one
buoyancy period, the flow will become strongly
anisotropic, but with the turbulence three-dimensional.  The stratified flow decays more slowly in time than does HIT
with identical initialisation.  Eventually, the flow becomes
quasi-two-dimensional, and the decay rate significantly increases.  The
foregoing summarises some of the results of \citet{debk19} and references
cited therein.

Here we consider simulations of initially isotropic SST similar to those
reported in the paper just cited, but with several important differences.
First, we add PSMLs oriented in the vertical and
spanwise directions, with the vertical axis anti-parallel to the
gravitational force and the active scalar maintained with a constant-in-time
uniform stabilising density gradient.  Second, the kinematic viscosity is
reduced by an order of magnitude relative to that in \citet{debk19} and
hyperviscous/hyperdiffusive terms are added such that the hyperviscosity and
hyperdiffisivity are equal, and the molecular Prandtl number $Pr = 0.7$.  The reason for doing this is to minimise the diffusive effects that are known to cause the mixing layer statistics to depart from self-similarity in the unstratified case and thereby, it is assumed, make the interpretation the effect of buoyancy in current simulations more straightforward.

The hyperviscosity model \citep[e.g.][]{lindborg06a} and other simple dissipation
models \citep[e.g.][]{diamessis05} have been shown to be effective for
studying SST at high Reynolds number and low Froude number when the details of
the small-scale statistics are not of interest; furthermore \citet{lalescu13} provide
theoretical and empirical evidence that such models are not expected to affect
dynamics in the inertial range in unstratified turbulence.  \citet{watanabe16}
show this to be the case for a stably stratified wake with $Pr=1$.  We note,
though, that recent studies of SST at higher $Pr$ suggest caution when using a
dissipation model for SST with $Pr>1$ because of a reverse buoyancy flux at small scales \citep{okino19,legaspi20,riley23,bragg24a,bhattacharjee26}.

This paper is organised as follows.  Theoretical background is provided in the
next section, followed by details of the simulations in
\S\ref{sec:simulations}.
Statistics of the velocity, active scalar, and passive scalars are in
\S\ref{sec:scalar}.  modelling approaches are considered in
\S\ref{sec:modelling}, followed by conclusions.

\section{Theoretical Background}
\label{sec:theory}
The simulations are solutions to the
non-divergent Navier-Stokes equations with the non-hydrostatic Boussinesq approximation.
The total density and pressure fields can be written
\begin{equation}
  \rho_t = \rho_0 + \frac{d \overline{\rho}}{d z} z + \rho
\end{equation}
\begin{equation}
  p_t = p_0 + p
\end{equation}
where $\rho_0$ is the (constant) background density
field and $d \overline{\rho} / d z$ is the (constant) background density
stratification. The notation $\overline{\left( \ \right)}$ indicates a planar average in
the horizontal.
The reference pressure, $p_0$, is in hydrostatic balance with the background
density field, i.e.,
\begin{equation}
  \frac{ d p_0}{d z} = -\rho_0 g z \ .
\end{equation}
The background density and stratification, along with the gravitational
acceleration $g$, combine to define the buoyancy frequency $N =
\left[-(g/\rho_0)  (d\overline{\rho}/dz)\right]^{1/2}$.

The Cartesian coordinate system is denoted $\vec{x}=\{x, y, z\}$ with
$z$ anti-parallel to the direction of gravity.
With this notation, and the unit vector in
the $z$-direction $\vec{e}_z$, the governing equations are
\begin{equation}
  {\nabla} \cdot \vec{{u}} = 0
\end{equation}
\begin{equation}
  \frac{\partial \vec{{u}}}{\partial {t}} +
  \vec{{u}} \cdot {\nabla} \vec{{u}} =
  -\frac{ {\nabla} {p}}{\rho_{o}}
  - \vec{e}_{z} \frac{g}{\rho_{o}} {{\rho}}
  + \nu {\nabla}^{2} \vec{{u}}
  + \nu_h {\nabla}^{4} \vec{{u}}
\end{equation}
\begin{equation}
  \frac{\partial {\rho}}{\partial {t}}
  + \vec{{u}} \cdot {\nabla} {\rho}
  + w \frac{d \overline{\rho}}{dz}
  = D {\nabla}^{2} {\rho}
  + D_h {\nabla}^{4} {\rho}
\end{equation}
\begin{equation}
  \frac{\partial {\xi_i}}{\partial {t}}
  + \vec{{u}} \cdot {\nabla} {\xi_i}
  = D {\nabla}^{2} {\xi_i}
  + D_h {\nabla}^{4} {\xi_i} \ .
\end{equation}
Here $\nu$ and $\nu_h$ are the kinematic and hyperviscous viscosities and $D$
and $D_h$ are the molecular and hyperdiffusive diffusivities.  The Prandtl
number for all scalars is $Pr = \nu / D = 0.7$, and $\nu_h=D_h$.  The passive scalars $\xi_i$ are taken to be mixture fractions so
that $0 \le  \xi_i \le 1$.  One passive scalar, denoted $\xi_z$, is initialised
with a mixing layer in the $z$-direction as sketched in figure
\ref{fig:schematic}, and the second passive scalar $\xi_y$ is initialised
with a mixing layer in the $y$-direction.

Further definitions of notation include the scalar profile widths and  the dissipation rates of kinetic energy.  The mean profile of a scalar mixing layer in the $z$-direction in HIT is very
nearly $\overline{\xi}= [1 + \mathrm{erf}(z/\delta)]/2$
\citep[e.g.][]{debk00a}, which motivates the definition of the profile width
$\delta$ as the distance from the centreline to the $z$ location at which
$\overline{\xi}=0.92$.  The width of $\overline{\xi}_y$ and $\overline{\xi}_z$
are denoted $\delta_y$ and $\delta_z$, respectively.  The domain-averaged dissipation rate of kinetic energy is $\epsilon_k=\epsilon_{k2} +
\epsilon_{k4}$, where $\epsilon_{k2}$ and $\epsilon_{k4}$ are the viscous and
hyperviscous contributions, respectively.

In figure \ref{fig:schematic}, the flow is shown to evolve spatially as it
would in the laboratory world, assuming that stratification is somehow suddenly
imposed on the turbulent flow.  Our simulated flows, though, are temporally
evolving, which avoids the numerical errors associated with the inlet and
outlet boundary conditions; \citet[][chapter 5]{pope00} discusses temporally
evolving flows.  Nevertheless, we use the terminology that $x$ is the streamwise and $y$ is the transverse direction.  The vertical direction is $z$.  In the simulations, the flow starts as HIT with
the volume-averaged r.m.s.\ value of any of the velocity components decaying
as $u' \propto (t-t_0)^{-n}$ with $t_0$ the virtual origin.  At $t=t_0$, the
integral length is $L_0$ and the r.m.s.\ velocity is $u'_0$, which combine to
form the time scale $\tau_0=L_0/u'_0$, and the definition
\begin{equation}
t^*\equiv(t-t_0)/\tau_0 \ .
  \label{eq:time}
\end{equation}

If the diffusivities of all the scalars are the same, as they are here, then
the flow can be parameterised in terms of three independent dimensionless
groups, namely a Reynolds, a Froude, and a Prandtl number.  The Prandtl number
is defined above.  For comparison to the literature on stratified turbulence
and passive scalar mixing layers (PSMLs), we define multiple Reynolds and Froude
numbers using these length and velocity scales: $L_h$ is the average of the
two horizontal longitudinal integral length scales each computed as
recommended in appendix~E of \citet{comte71}; $L_t = u'^3/\epsilon_k$ is the
turbulent length scale; $\lambda$ is the transverse Taylor microscale; $u'_h$
is the average of the two horizontal r.m.s.\ velocities.  Then
\begin{alignat*}{3}
  \Rh=\frac{u'_h L_h}{\nu} \ \ \ \ \ &
  \Rey=\frac{u' \lambda}{\nu} \ \ \ \ \ &
  \Rt = \frac{u'^4}{\nu \epsilon_k} \\
  \Fh=2\pi \frac{u'_h}{N L_h} \ \ \ \ \ &
  \Ft=\frac{\epsilon_k}{N u'^2} \\
  \Rb=\Rh \Fh^2 \ \ \ \ \ &
  Gn=\frac{\epsilon_k}{\nu N^2} \ .
\end{alignat*}

Both $\Rb$ and $Gn$ are referred to in the literature as the ``buoyancy Reynolds number.'' The latter, which is also called the ``activity parameter,'' is defined by \citet{gibson80} and identified as the scale separation between the Ozmidov length scale $L_O$ and the Kolmogorov length scale $L_K$, specifically $Gn=(L_O/L_K)^{4/3}$, as used by \citet{gargett84}.  In contrast, $\Rb$ derives from a criterion based on the gradient Richardson number required for a stratified flow to become or remain turbulent \citep{riley03}.  If $\Rb$ is defined in terms of the length scale $L_t$ instead of $L_h$ then $\Rb = Gn$.  The time evolution of $Gn$ and $\Rb$ for a flow similar to, but at lower Reynolds number, than that studied here is plotted as figure 10 in \citet{debk19}.

\begin{table}
  \begin{center}
    \begin{tabular}{ l l r r r r r}
      &
      & \multicolumn{2}{c}{Unstratified} &
      & \multicolumn{2}{c}{Stratified} \\
      \cmidrule{3-4} \cmidrule{6-7}
      & & $t^*=1$ & $t^*=10$ && $t^*=1$ & $t^*=10$
      \\
      Integral Reynolds number & $\Rh$ & 23250 & 16148 && 21552 & 33658 \\ 
Taylor Reynolds number & $\Rey$ & 734 & 479 && 734 & 980 \\ 
Turbulent Reynolds number & $\Rt$ & 35892 & 15291 && 35892 & 63982 \\ 
Activity parameter & $\Gn$ &   &  && 1545 & 11 \\ 
Buoyancy Reynolds number & $\Rb$ &   &  && 99835 & 1383 \\ 
Horiz.\ Froude number & $\Fh$ &   &  && 2.15 & 0.20 \\ 
Turbulent Froude number & $\Ft$ &   &  && 0.207 & 0.013 \\ 
Horiz.\ domain size / integral length & $\mathcal{L}_h / L_h$ & 74.2 & 25.8 && 74.2 & 18.2 \\ 
Grid spacing / Kolm.\ length & $ \Delta / L_K$ & 28.5 & 9.0 && 28.5 & 8.3 \\ 
Buoyancy length / Grid spacing & $ L_b / \Delta $ &  &  && 118.8 & 45.6 \\ 
Horiz.\ domain size to vert.\ domain size & $\mathcal{L}_h/\mathcal{L}_v$ &  2.0 & &&  2.0 \\ 
Horiz.\ grid points & $N_x$,$N_y$ & 4096 & && 4096 \\ 
Vertical.\ grid points & $N_z$ & 2048 & && 2048 \\ 
Resolved diss.\ rate & $\epsilon_2/\epsilon$ & 0.26 & 0.45 && 0.26 & 0.47 \\ 

    \end{tabular}
    \caption{\label{tbl:one} Simulation parameters and metrics at two
      times for each simulation.}
  \end{center}
\end{table}

\section{Overview of Simulations}
\label{sec:overview}
\label{sec:simulations}

A stratified and an unstratified simulation are considered for this research.
The stratified case is very similar to Case III in \citet{debk19}, except that
here the Reynolds number is higher and hyperviscosity and hyperdiffusivity
are used.  The unstratified case is the same as the stratified case except the density field $\rho$ is zero and $d\overline{\rho}/dz=0$.  In other words, the same algorithm is used for both simulations.  Various simulation parameters and
statistics, defined below in this section, are in table \ref{tbl:one}.

The simulation methodology is very similar to that of \citet{debk19} except
that dealiasing is done with a combination of phase shifting and the
$2\sqrt{2}/3$-method of truncation as proposed by \citet{rogallo81}, and
time stepping is done with a third-order Runge-Kutta schema.  In summary,
derivatives, the pressure gradient, and addition are computed in Fourier
space while multiplication is done in real space.  The diffusion terms are
computed with integrating factors.

\begin{figure}
  \includegraphics{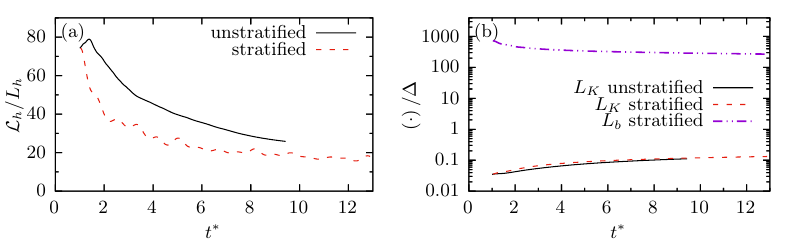}
  \caption{\label{fig:resolution} Measures of the large and small-scale resolution
    in the simulations.}
\end{figure}

Sufficient large-scale resolution is essential in simulations of decaying
turbulence.  This can be expected because the decay rate depends on the shape
of the kinetic energy spectrum at large length scales \citep{saffman67}, with
analysis extended to stratified flows by
\citet{davidson10}. The resolution is expressed in terms of the horizontal and vertical dimensions of the domain denoted ${\mathcal L_h}$  and ${\mathcal L_v}$, respectively.  \citet{debk98b} empirically determine the criterion
${\mathcal L_h}/\Lh \ge 20$ in HIT for the decay rate to be consistent with laboratory
experiments. This criterion is shown to be met in figure
\ref{fig:resolution}(a).

Measures of small-scale resolution of the simulations are given in figure
\ref{fig:resolution}(b) in terms of the grid spacing $\Delta$, the Kolmogorov
length scale $L_K=(\nu^3/\epsilon_k)^{1/4}$, and the buoyancy length scale
$L_b=2\pi u_h/N$, with the factor of $2\pi$ retained to be consistent with
\citet{waite11} and others.  Criteria for small-scale resolution of
homogeneous flows has been studied extensively for the purpose of computing
quantities such as enstrophy and moments of local $\epsilon_k$
\citep[e.g.][]{yeung18}. Comparable resolution is likely required for studies
of extreme events in stratified turbulence, such as that done by
\citet{petropoulos24}. For current purposes, though, we rely on the analysis
of \citet{lalescu13} supported by empirical evidence
\citep[e.g.][]{diamessis05,lindborg06a} to conclude that the resolution indicated
by figure \ref{fig:resolution}(b) coupled with the hyperviscous model is
sufficient for the statistics reported in this paper.  Importantly,
$L_b/\Delta$ greatly exceeds the criterion established by \citet{waite11} that
the buoyancy scale be resolved.

\begin{figure}
  \includegraphics{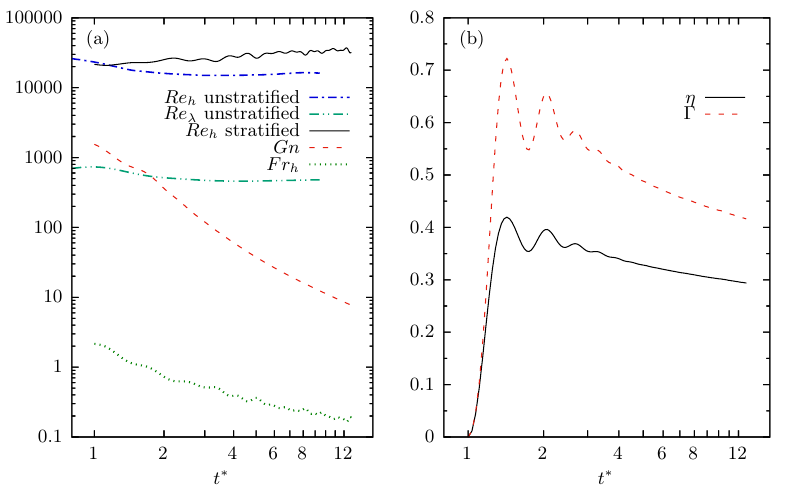}
  \caption{\label{fig:stats} Statistics of the velocity fields and of
    mixing of the active scalar.}
\end{figure}
Some statistics of the velocity fields are provided in figure
\ref{fig:stats}(a).  Considering the unstratified case first, $\Rh>15,000$ and
$\Rey>450$ for the duration of the simulation, which is well above the
threshold for inertial-convective and inertial-diffusive ranges to exist at
$Pr=O(1)$, as reviewed by \citet{shete20}.  Note that $\Rey$ for the stratified
case is not included on the figure because the flow is strongly anisotropic so that assumptions in the definition of the Taylor microscale are not met,
but values at two times are in table \ref{tbl:one} for reference.  Turning now to the
stratified case, by $t^*=2$, $\Fh$ is less than unity, which is indicative
of strong stratification for a Froude number defined this way \citep{debk19}.
At the same time, $Gn$ is large and remains so for the duration of the stratified
simulation.  So the simulation is strongly stratified and strongly
turbulent.

\label{sec:stiringVmixing}
The process colloquially known as mixing can be separated into macroscopic reversible ``stirring'' and microscopic irreversible ``mixing'' \citep{eckart48}.  Two measures of mixing of the active scalar are shown in figure \ref{fig:stats}b.  Molecular diffusion reduces the variance of the active scalar, which is related to the available potential energy of the flow.  The mixing parameter $\Gamma=\epsilon_p/\epsilon_k$ and mixing
efficiency $\eta=\epsilon_p/(\epsilon_p+\epsilon_k)$ are consistent with those
reported in the literature for a variety of strongly stratified flows with
$Pr=O(1)$ \citep[e.g.][]{augier12,salehipour15,maffioli16b}.  The co-variation of $\Gamma$ and $Fr_t$, with both falling over time (Figure~\ref{fig:stats}b, Table~\ref{tbl:one}), also matches the parametric dependence of $\Gamma$ on $Fr_t$ for the stably-stratified regime \citep{maffioli16b,bragg24b}.

\begin{figure}
  \includegraphics{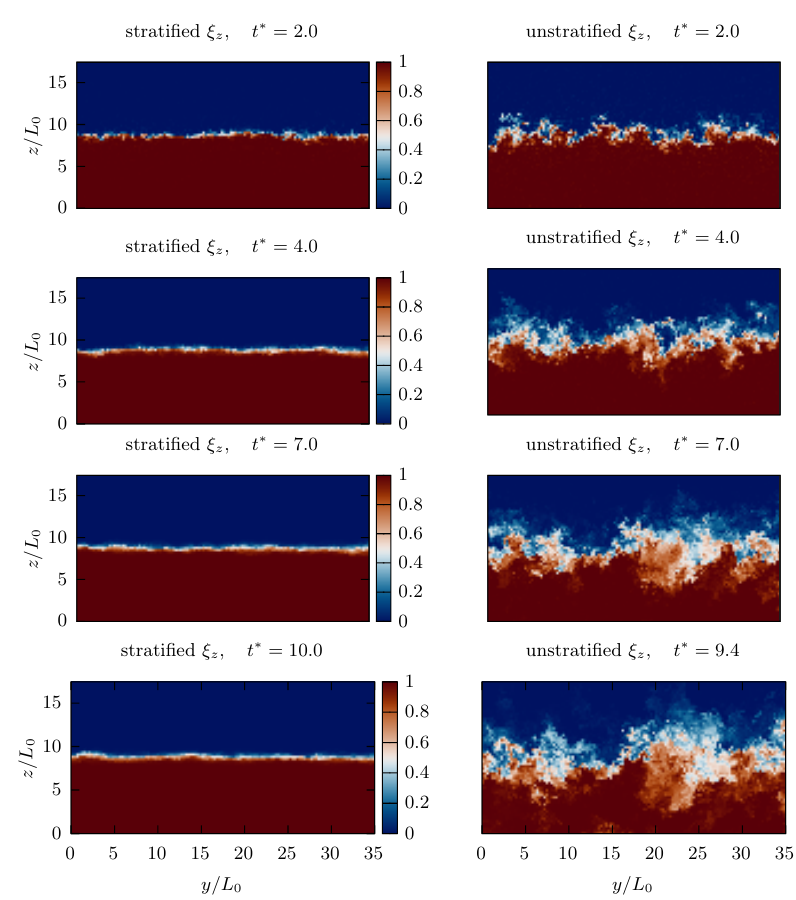}
    \centering
  \caption{Slices of $\xi_z$ through the vertical centrelines for the stratified (left) and unstratified (right) cases.  Only half of the domain in each direction is shown.  Note that slices of $\xi_y$ and $\xi_z$ are qualitatively similar for the unstratified case and that only the latter are shown.
  \label{fig:xiZSlice}}
\end{figure}

\begin{figure}
  \includegraphics{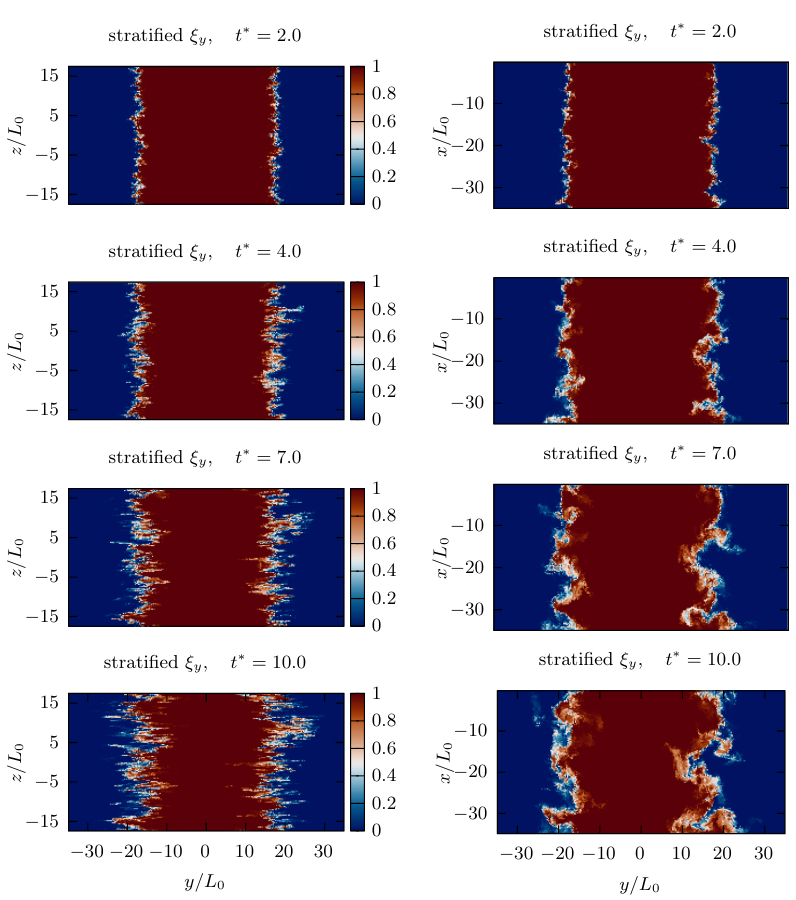}
  \centering
  \caption{Slices of $\xi_y$ through the
    vertical (left) and transverse (right) centrelines for the stratified case.  Only half the domain is shown in the
  transverse direction.  Note that each mixing layer is two-sided to enable using periodic boundary conditions.
  \label{fig:xiYSlice}}
\end{figure}

Some understanding of the evolution of the passive scalars is provided by figures
\ref{fig:xiZSlice} and \ref{fig:xiYSlice}.  These slices give an overview,
while the details of the scalar statistics are in \S\ref{sec:scalar}.  In the
first of these figures are shown slices of $\xi_z$, which is the layer having a mean gradient in the vertical.  Images for the stratified and unstratified cases are shown.  In the unstratified case, the effect of stirring generating scalar intermittency is evident, as is the effect of mixing that produces broad regions of scalar with values near $\xi_z\approx 0.5$.  In the stratified cases, in contrast, stirring is curtailed by buoyancy and the layer spreads very little in time.

In figure \ref{fig:xiYSlice}, horizontal and vertical slices from the stratified case are shown for $\xi_y$, which is the layer having a mean gradient  oriented in the transverse direction.  Significant spreading is observed in the horizontal, but the character of the intermittency is observably different from that of unstratified layer in figure \ref{fig:xiZSlice}.  Slices on the $yz$ plane show that very fine layers form in the $z$-direction so that, while the layer is homogeneous in both $x$ and $z$, the structure is much different in the two directions.

\section{Passive scalar statistics}
\label{sec:scalar}
\subsection{Spreading rate}
\begin{figure}
  \includegraphics[width=\textwidth]{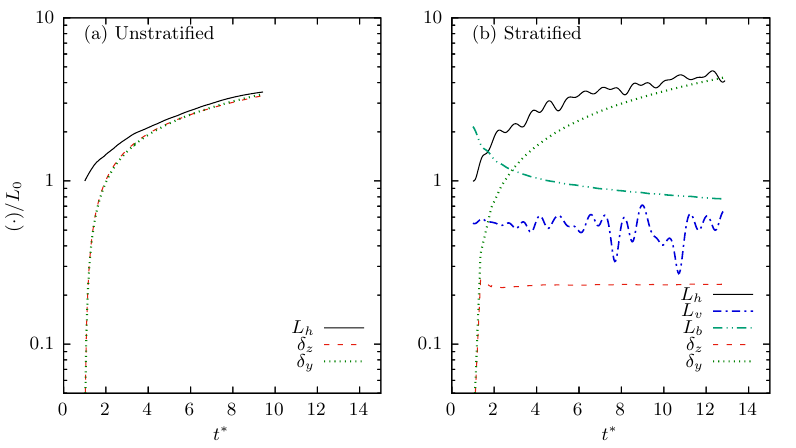}
  \caption{Width of scalar mixing layers $\delta_y$ and $\delta_z$ along with various other
    length scales.  $\delta_y$ and $\delta_z$ are the profile widths for $\overline{\xi}_y$ and $\overline{\xi}_z$, respectively.
    \label{fig:lengths}}
\end{figure}

One of the most basic measures of the scalar mixing layer is the rate at which it
spreads in time.  These rates are shown in figure \ref{fig:lengths}, along
with length scales of the velocity fields for comparison.  For the
unstratified case (figure.~\ref{fig:lengths}a), the layer is expected to be
approximately self-similar so that the layer width is proportional to the
integral length scale of the velocity field.  This is the case when the virtual origins of the
velocity and scalar fields are matched \citep{debk00a}.  Otherwise, as time
evolves, the scalar
width $\delta$ approaches the velocity length scale \citep[e.g.][]{ma86}.
This latter behaviour is observed in figure \ref{fig:lengths}a.
Similar
behaviour is observed in figure~\ref{fig:lengths}b
for the stratified case with the layer oriented in the
transverse direction; $\delta_y$ approaches $L_h$ as time advances.  In fact, the growth rate of this
layer is very similar to that of the unstratified case, but the velocity length
scale grows faster so that $\delta_y$ is still approaching $L_h$ even at late time.

For the stratified case with the layer oriented in the vertical, the layer depth,
$\delta_z$, quickly grows to some thickness and then approximately stops
growing. The time at which it approximately stops growing is less than one buoyancy period; one buoyancy period after gravity is activated corresponds to $t^*=3$.  This inhibition in growth is due to two related effects.  The first is that the stable density stratification causes the conversion of the vertical component of the kinetic energy into potential energy, thus impeding the vertical growth of the layer.  The second is that, by one buoyancy period, the flowfield has become somewhat internal wave-like, and internal waves are inefficient at mass transport.  This wave-like behavior can be seen in the oscillations of the vertical component of the velocity in Figure~2 of \cite{debk19}.  Two vertical length scales of the velocity field are
included in figure~\ref{fig:lengths}b.  $L_v$ is the vertical integral length scale of the horizontal motion, which is
the average of those scales for $u$ and $v$.  There is a lot of scatter in
this statistic, but it appears to oscillate about a constant and so is consistent
with the invariance of $\delta_z$ following its initial transient.  The second vertical velocity scale
in the figure is the buoyancy length scale $L_b$.  It decreases in time
throughout the simulation, unlike $\delta_z$.  The results are hardly
definitive, but they suggest that $L_v$ and not $L_b$ is the length scale that controls
$\delta_z$.   After the initial transient, however, the distinction becomes moot because it is expected that $L_v$ will adjust so that the vertical Froude number $Fr_v=2 \pi u'_h /(N L_v) = O(1)$ \citep{billant01}, which is shown to be the case for a similar flow \citep[][figure 9]{debk19}. After $Fr_v$ has become approximately constant in time, $L_v \propto L_b$, and both describe the limit on $\delta_z$.  Note that suppression of the growth of the layer in the vertical, as exhibited by the behaviour of $\delta_z$, is qualitatively similar to the prediction of \cite{csanady64}.

\subsection{Moment and flux profiles of the transverse layer}
\begin{figure}
  \includegraphics[width=\textwidth]{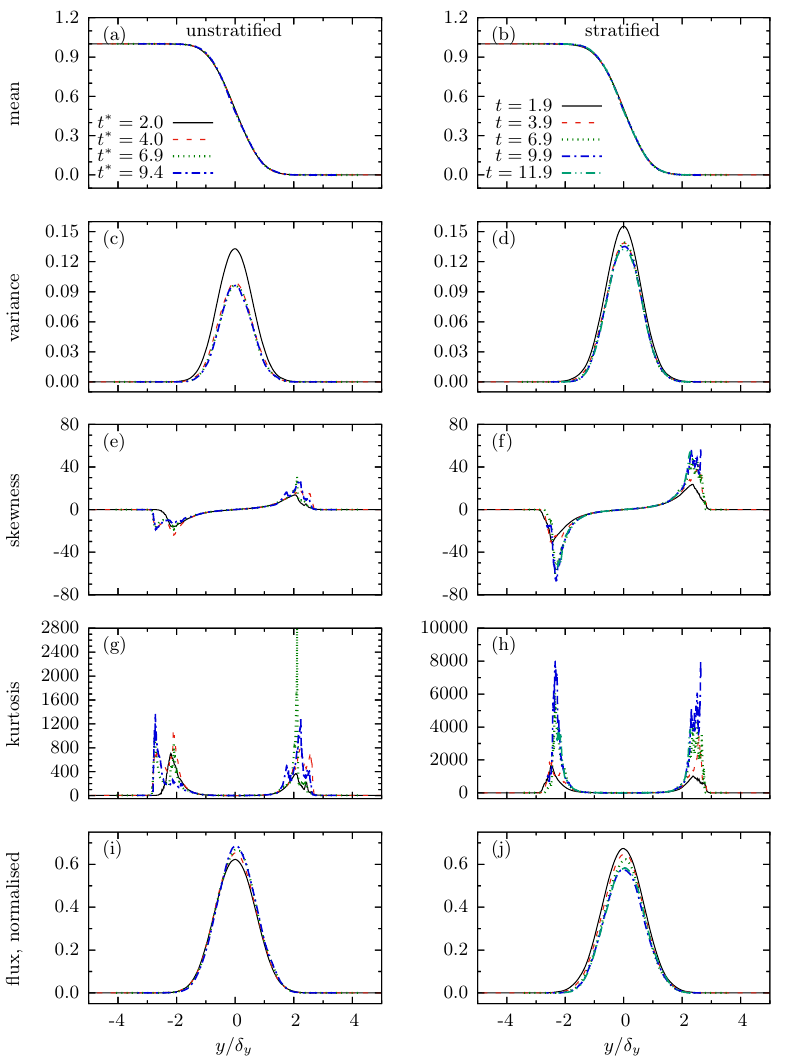}
  \caption{\label{fig:z1Profs} Moment and flux profiles for the layer $\xi_y$ over part of the spatial domain.}
\end{figure}

Figures \ref{fig:xiZSlice} and \ref{fig:lengths}b indicate that the layer oriented with mean scalar gradient in
the $z$-direction does not spread very much after early time in the stratified case.  So for now let us focus on the layer with mean gradient in the transverse direction
and return to the vertical layer later in this section.  Moment and flux profiles
of the transverse layer are shown in figure \ref{fig:z1Profs}.

Beginning with the unstratified case, it is observed that the layer is very nearly an error function with Gaussian variance profile.  After an initial transient, the peak variance converges to approximately 0.09, which corresponds to a peak intensity of 0.3.  This is in excellent agreement with the theoretical prediction of \citet{libby75} based on similarity analysis, i.e., assuming high $\Pe$, but is higher than measured values at moderate $\Pe$ \citep{larue81a,ma86}, which is consistent with the correction for moderate $\Pe$ developed by \citet{debk00a}.  The peak skewness and kurtosis are somewhat higher than those reported in the references just cited, and this is consistent with the current simulations having higher $\Pe$.  The flux normalised by the r.m.s.\ of the $v$ velocity and the peak intensity does not collapse perfectly but is very close to the published results.  Taking these profiles together, we conclude that the unstratified case has the characteristics of a PSML at high Reynolds and P\'eclet numbers expected from the literature.

Turning now to the stratified case, it is observed that the statistics are qualitatively similar to those of the unstratified case with the principal difference being that the scalar fluctuations are more intense in the stratified case.  This is consistent with the large-scale intermittency in $\xi_y$ at the edges of the stratified layer (figure \ref{fig:xiYSlice}, right hand column) compared with that in $\xi_z$ the unstratified case (figure \ref{fig:xiZSlice}, right hand column), where slices of $\xi_z$ in the unstratified case are also representative of those of $\xi_y$ due to isotropy. The larger fluctuations result in higher peak variance, skewness, and kurtosis.  The normalised flux is slightly lower in the stratified case; the flux will be considered in detail in \S\ref{sec:modelling}.

\subsection{Stirring and mixing of the transverse layer}
Stirring and mixing are defined in \S\ref{sec:stiringVmixing}.  The latter is the term we use for destruction of the active scalar variance by molecular diffusion, c.f. figure \ref{fig:stats}, and we continue with that definition for the passive scalar.  Figures \ref{fig:xiZSlice} and \ref{fig:xiYSlice} suggest that in the stratified case the flow is less effective at mixing the passive scalar near the center of the layer, as there are somewhat sharply defined fingers of high $\xi$ extending into low $\xi$ whereas these are diffused in the unstratified case.  We can quantify the unmixedness with the parameter
\begin{equation}
0 \le \Xi = \frac{\overline{\xi^{\prime 2}}}{\overline{\xi}(1-\overline\xi)} \le 1 \ ,
\label{eq:unmix}
\end{equation}
where $\overline{\left(\cdot\right)}$ indicates a planar mean and $\left( \cdot \right)^\prime$ is the fluctuation about that mean \citep{danckwerts52,dimotakis90}.
\begin{figure}
  \includegraphics[width=\textwidth]{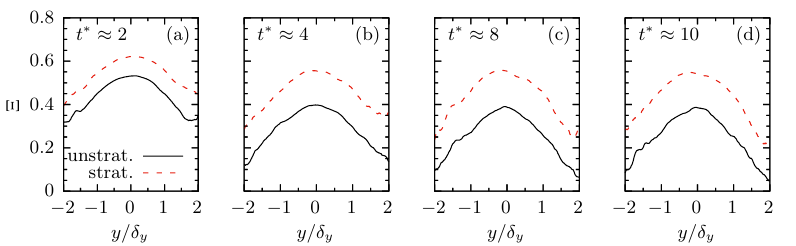}
  \caption{\label{fig:unmix} Unmixedness parameter at selected times.}
\end{figure}
The unmixedness parameter is plotted as figure \ref{fig:unmix} from which several observations can be made.  First, in the stratified case the flow is significantly less mixed than the unstratified case, as we hypothesised from observing slices through the layer.  Second, the unmixedness becomes approximately constant in time, which is consistent with the mixing being self-similar.

\begin{figure}
\centering
  \includegraphics{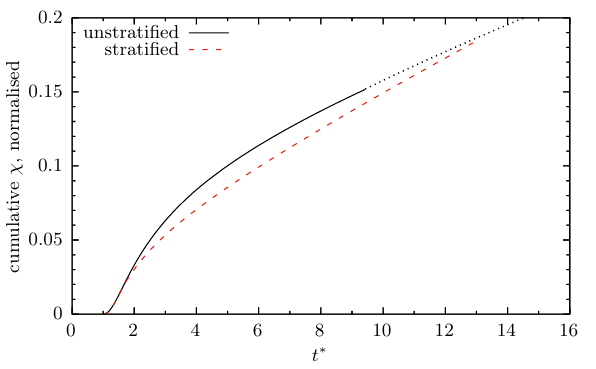}
  \caption{\label{fig:chi} Cumulative integral of the dissipation rate of the variance of $\xi_y$ normalised by the initial value of that variance.  The dotted line is extrapolated.}
\end{figure}

While $\Xi$ is informative about the unmixedness within several layer thicknesses of the centreline, it is numerically intractable at large values of $y/\delta$ because both the numerator and the denominator approach zero. A direct measure of mixing in the entire domain is the domain average dissipation rate of the passive scalar variance $\chi = \chi_{k2}+\chi_{k4}$ where $\chi_{k2}$ and $\chi_{k4}$ are the diffusive and hyperdiffusive contributions, respectively.  The cumulative time integral of $\chi$ is plotted in figure \ref{fig:chi}.
At early times, the unstratified case mixes somewhat
faster, but eventually $\chi$ for the stratified case exceeds that for the unstratified case at
the last time for which both are available. This suggests that the cumulative mixing in
the stratified case would exceed that of the unstratified case at late times if the large
scale resolution constraint did not limit the duration of the simulations.  These results can be explained by recalling that mixing follows stirring.  At early times the effect of stable stratification is to suppress the vertical velocity, as a significant mount of kinetic energy is converted into potential energy \citep{debk19}.  Therefore in the stratified case less kinetic energy, and hence velocity, is available to stir the scalar field.  The kinetic energy decays more slowly in the stratified case, however \citep{debk19}, so that at later times the kinetic energy in the stratified case exceeds that in the unstratified case.  Furthermore, recall from figure \ref{fig:lengths} and table \ref{tbl:one} that $L_h$ for the stratified case eventually grows larger than for the unstratified case.  These results both indicated that eventually more mixing will occur for the stratified case.



\subsection{Mixing in the vertical in the stratified case}
\label{sec:verticalMixing}
\begin{figure}
\includegraphics[width=\textwidth]{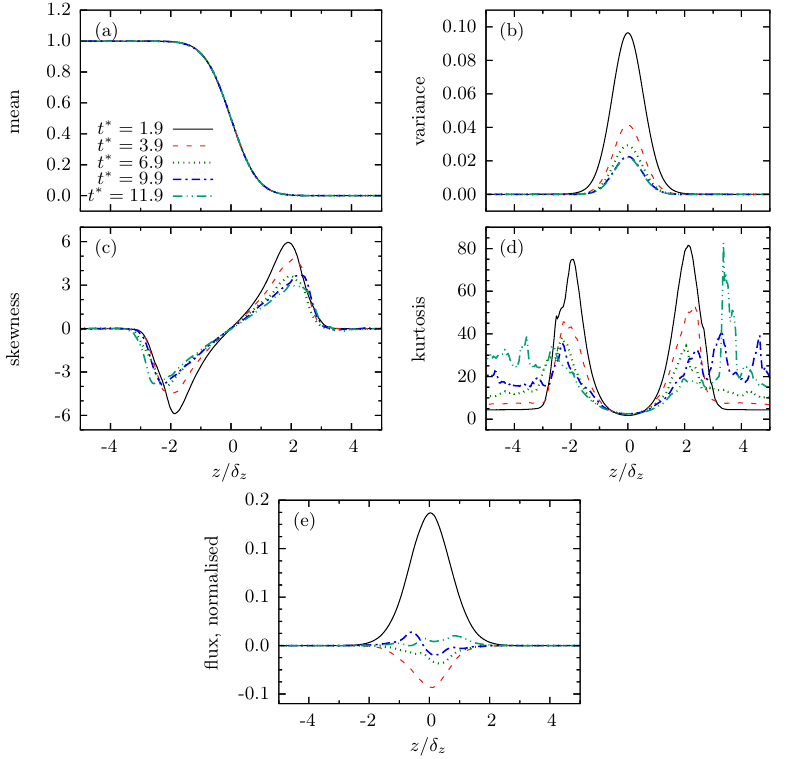}
\caption{\label{fig:z0Profs} Moment and flux profiles for the vertical layer $\xi_z$.}
\end{figure}

We have concluded from figures \ref{fig:xiZSlice} and \ref{fig:lengths}b that the layer with an initially vertical gradient in the passive scalar grows at very early times and then almost stops growing as stirring is curtailed by stratification.  For completeness, though, the passive scalar moment and flux profiles of $\xi_z$ are given in figure \ref{fig:z0Profs}.
With little stirring of unmixed scalar from the edges to the middle of the layer, the variance and higher moments of the scalar fluctuations decay in time. The peak skewness and kurtosis are small even compared with those for the unstratified case.

\begin{figure}
\includegraphics[width=\textwidth]{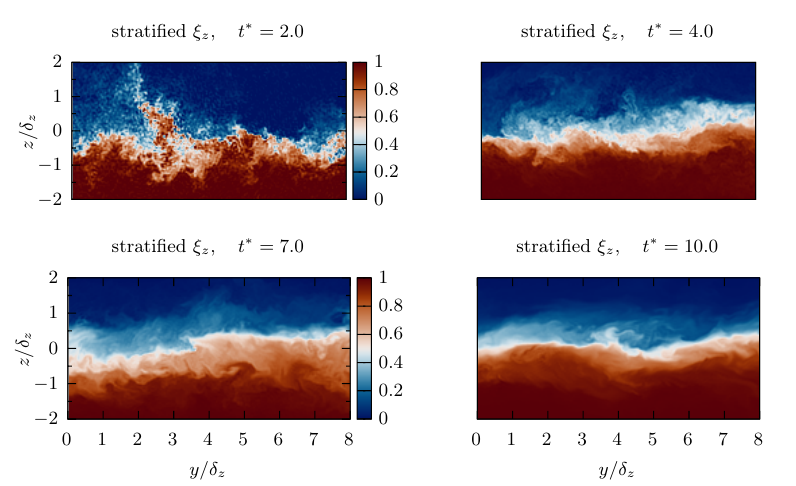}
\caption{\label{fig:xiZaSlice} Slices of $\xi_z$ through the centreline at the full resolution of the simulations so that only a small part of the domain is shown.  The dimensions are scaled by $\delta_z$ for comparison with figure \ref{fig:z0Profs}; $\delta_z / L_0 \approx 0.24$ for all times shown here.}
\end{figure}

The vertical flux merits some additional consideration.  At early time it is positive, indicating a conversion of kinetic energy into potential energy \citep{debk19}; but the sign reverses after several buoyancy periods, and then the flux goes to approximately zero by late time.  Note that $t^*=3$ corresponds to one buoyancy period after the buoyancy force is initialised.  No reverse flux is evident in the mean profile by zooming in on figure \ref{fig:z0Profs}a.  Indeed, the steepening of the passive scalar profile would indicate impossible unmixing, and $\delta_z$ does grow monotonically, although by a very small amount, as it must owing to molecular diffusion.  The turbulent flux, though, is a stirring process, and indications of negative flux are evident in figure \ref{fig:xiZSlice} by noting the details of the red-blue interface.  The interface region for the stratified cases is reproduced at higher vertical resolution as figure \ref{fig:xiZaSlice}.
At $t^*=2.0$ there is a tendril of red near $y/\delta=2.0$ that is largely gone by $t*=4.0$ and completely gone by $t*=7.0$.  Since one buoyancy period after gravity is initiated corresponds to $t^*=3$, this tendril formed before buoyancy curtailed stirring in the vertical.  While its fate cannot be proven without tracking the fluid elements, its disappearance is consistent with a negative flux sweeping the tendril back toward the side of the layer with $\overline{\xi}_z=1$.  Also, there is internal wave-like behaviour expected in this flow, analogous to the oscillating buoyancy flux observed by \citet[][figure 6]{debk19}, which will tend to cause oscillations in the vertical flux of the passive scalar.

\section{Modelling the transverse scalar flux}
\label{sec:modelling}
A variety of modelling techniques for passive scalars are reported in the literature, and their applicability depends on the information available as model inputs and the fidelity required of the model.  For Reynolds Averaged Navier Stokes simulations, models with a turbulent Prandtl number $Pr_T$ are used with $Pr_T=0.85$ being suitable in boundary layers and lower values for free-shear flows \citep{kays94,kays12}.  In large-eddy simulation, the analogous quantity is a subgrid Prandtl number, with appropriate values found to be in the range 0.3 to 0.5 \citep{mason90,moin91}.  More recently, p.d.f.\ methods have been developed \citep[c.f.][]{celis15}; these are attractive, particularly for flows with high $\Pe$ or chemical reactions, because the advective and reaction terms are closed.  To use any of these methods to model mixing effectively in a stratified flow, it is necessary to accurately simulate the velocity field, which is a topic of current research \citep[e.g.][]{li22}. This is especially true here, as our turbulent velocity field is evolving \citep{debk19}, as opposed, for example,  to the statistically steady flow field of \cite{lindborgfed09}.

Outside of simulations, it may be necessary to estimate mixing given limited information about the velocity and scalar fields.  This is our starting point here.  We consider several eddy diffusivity models that differ by whether the scalar profile is known or not.  We limit the study to the transverse scalar flux for the cases with the mean scalar gradient in the transverse direction. The results in \S\ref{sec:verticalMixing} show very little mixing in the vertical, due to the inhibition of the vertical component of the velocity by the stable density field.  Therefore accurately modelling the vertical scalar flux would also require accurate modelling of this component of the velocity field as affected by the stable stratification.

\subsection{Known scalar profile}
The eddy diffusivity is $D_T$ defined by
\begin{equation}
F_y \equiv \overline{v \xi_y} = - D_T \, \dzy\ .
\label{eq:eddyDiffusivity}
\end{equation}
where $F_y$ is the flux and the notation $\dzy \equiv \partial\overline{\xi}_y/\partial y$ has been introduced for compactness.
Applying mixing-length theory \citep{taylor1915}, $D_T$  can be modelled with a velocity ${\cal U}$ and a length ${\cal L}$:
\begin{equation}
D_T \propto \, \calU \calL \ .
\end{equation}
Given that $\dzy$ is assumed to be available for this approach, $\delta_y$ is also available.  Assuming $\calL \propto \delta_y$ and $\calU=v'$, the r.m.s.\ of $v$, the model for flux becomes
\begin{equation}
F_{1} = - c_1 \delta_y v' \dzy \ ,
\label{eq:F1}
\end{equation}
with $c_1$ expected to be $O(1)$.

\begin{figure}
  \includegraphics{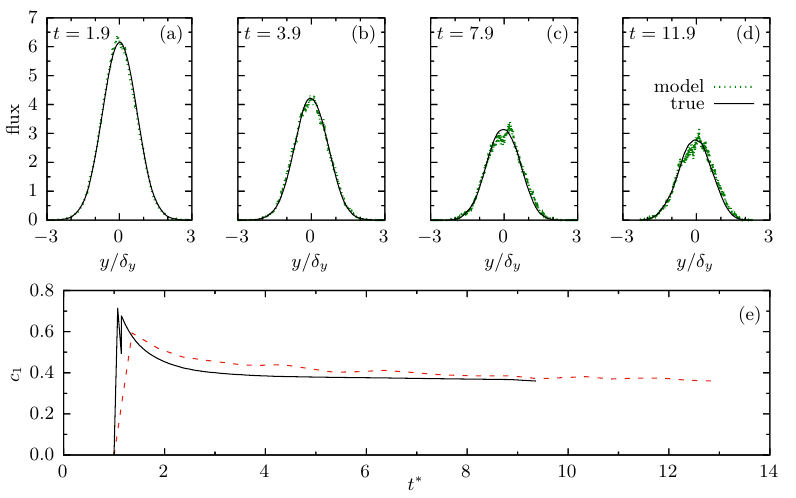}
    \centering
  \caption{\label{fig:turbDiff2} True and model $F_{1}$ for the stratified case are shown in panels (a), (b), (c), and (d).  The constant $c_1$ for the unstratified (solid line) and stratified (dashed) line is in panel (e).  The constant is computed by minimising the sum of the squares of the differences between the true and model fluxes.}
\end{figure}

The true and modelled fluxes are plotted for the stratified case as figure \ref{fig:turbDiff2}, from which it is evident that the model is excellent and the constant $c_1$ is, indeed, order one after an initial transient.
From the results in \citet{debk00a}, this model will be effective for the unstratified case and so these results are not shown except for $c_1$, which is included in
panel (e) of the figure. After the initial transient, when the layer grows very rapidly, the model constant $c_1 \approx 0.4$ for the stratified and unstratified cases, and slowly decreases in time as the $\Pe$ number decreases in time because of finite molecular diffusivity.  Inherent in the value of $c_1$ is the definition of the scalar length scale $\delta$.  If a different definition of $\delta$ is used, such as half width at half height, then the value of $c_1$ will change accordingly, but it should not be affected by whether or not the flow is stratified.

\subsection{Assumed scalar profile}
\begin{figure}
  \includegraphics{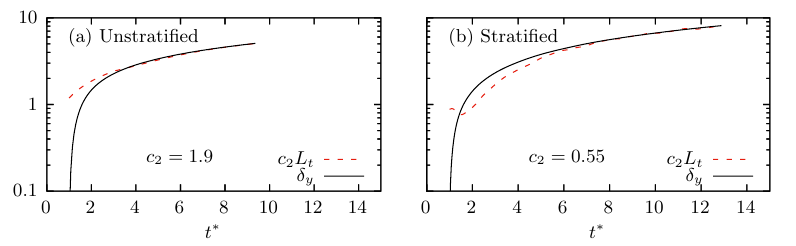}
    \centering
  \caption{\label{fig:deltaModel} The scalar profile width and the model for it given by the right hand side of \eqref{eq:deltaModel}.}
  \end{figure}

If the scalar profile is not known then it may be estimated by observing from figure \ref{fig:z1Profs} that it is very nearly an error function.  What is required is a surrogate for $\delta_y$.  A classical approach to modelling a turbulent length scale is used in our definition of $\Ft$, namely
\begin{equation}
L_t = v'^3/\epsilon_k \ .
\end{equation}
In HIT, $L_h/L_t$ approaches a constant at high Reynolds number \citep[figure 6.24]{pope00}, and $\delta_y/L_h$ is constant in the self-similar case \citep{debk00a}, c.f.~figure \ref{fig:lengths}.  Therefore, we can expect an effective model for the scalar profile width to be:
\begin{equation}
\delta_y = c_2 L_t
\label{eq:deltaModel}
\end{equation}
with $c_2$ an $O(1)$ constant.  Indeed, figure \ref{fig:deltaModel}a shows the model to be excellent after the transient at early time.

For the stratified case, there are several reasons to doubt the approximation in \eqref{eq:deltaModel}.  One is that figure \ref{fig:lengths} shows that $\delta_y$ approaches $L_h$ in time but does not support the approximation $\delta_y \propto L_h$ over the time range of the simulation.  A second is that the ``constant'' of proportionality between $L_t$ and $L_h$ is known to vary significantly with horizontal Froude number \citep{hebert06b,debk19}, which is important because the that Froude number decreases in a decaying stratified flow as the length scale grows and the velocity scale decreases in time.
Nevertheless, we proceed by computing $c_2$ given $L_t$ from the simulations and $c_1=0.4$  The true width $\delta_y$ and the model width $c_2 L_t$ are plotted in as figure \ref{fig:deltaModel}b.  The value of $c_2=0.55$ is lower than for the unstratified case but still $O(1)$.

\begin{figure}
  \includegraphics{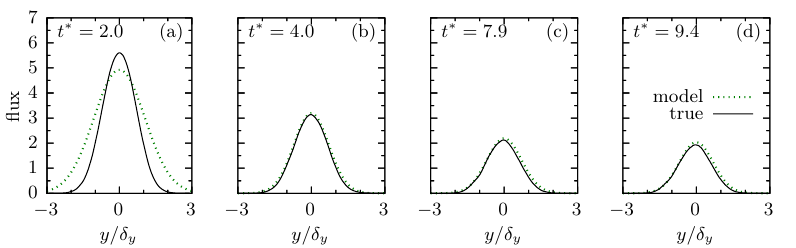}
    \centering
  \caption{\label{fig:F2_i} True and model $F_{2}$ flux profiles for the unstratified case with $c_1=0.4$ and $c_{2}=1.9$.}
\end{figure}

Writing the mean scalar profile in terms of $\delta_y = c_2 L_t$ yields
\begin{align}
\overline{\xi}_y &= \frac{1}{2}\left[1 - \mathrm{erf}\left(\frac{y}{c_2 L_t} \right)\right] \ ,\\
\partial \overline{\xi}_{y,y} &= -\frac{1}{c_2 L_t \sqrt{\pi}} \exp\left(-\frac{y^2}{(c_2  L_t)^2} \right) \ , \\
F_{2} &= \frac{c_1 \, v'}{\sqrt{\pi}} \exp\left(-\frac{y^2}{(c_2  L_t)^2} \right)
\end{align}
with $F_{2}$ being the model of the flux assuming a mean scalar profile.  This model is plotted for the unstratified case as figure~\ref{fig:F2_i}.  At early times, the model flux profile is too wide, which is consistent with the passive scalar not yet being in quasi-equilibrium with the velocity.  The excellent model performance at later times reflects that the model coefficients are, in effect, tuned to give the height and width if the scalar is in quasi-equilibrium with the velocity.

\begin{figure}
  \includegraphics{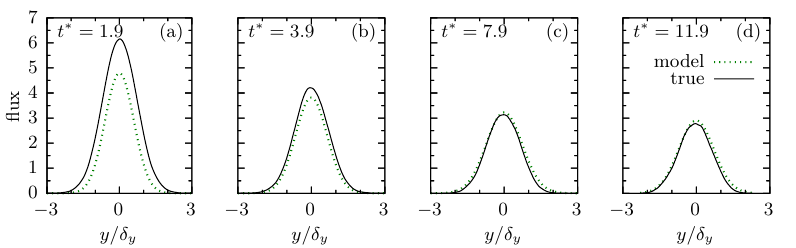}
    \centering
  \caption{\label{fig:F2_2} True and model $F_{2}$ flux profiles for the stratified case with $c_1=0.4$ and $c_{2}=0.56$.}
\end{figure}

For the stratified case, the model is plotted as figure~\ref{fig:F2_2}.  Again, the model is excellent at late times for the same reason it is for the unstratified case.  Interestingly the model under-predicts both the height and the width of the flux profile at early time.  If $c_{2}$ for the unstratified case is used then the model significantly over-predicts the width of the flux profile.  This simple model does not account for the scalar time scale being out of quasi-equilibrium with the velocity time scale at early time, and it does not account the velocity time scale changing due to the effects of stratification imposed at $t^*=1.0$.   Nevertheless, it does give good predictions of the flux at late time.

\section{Conclusions}
\label{sec:conclusions}
Two passive scalar mixing layers (PSMLs) with mean gradients oriented in the vertical and transverse directions, respectively, are a model for a plume that is much larger than the velocity length scale for the flow. The behaviour of PSMLs in homogeneous isotropic turbulence (HIT) has been reported extensively in the literature, from which it might be concluded for modelling purposes that the scalar statistics will approximate self-similarity in time if the effects of molecular diffusion are negligible.  Therefore, to study passive scalar mixing in stratified turbulence, two highly resolved large-eddy simulations were developed in which the Reynolds and P\'eclet numbers are high and the duration in time is sufficient for the scalars to come into quasi-equilibrium with the velocity field.  One simulation is homogeneous and isotropic in power-law decay.  The second is identical to the first except that gravity acts on a uniform stabilising density gradient so that the HIT becomes strongly stratified after about one buoyancy period, which corresponds to two large-eddy times.

The simulations show several important characteristics of passive scalar mixing in stratified environments.  The first is that, with the scalar gradient oriented in the transverse direction, passive scalar mixing rates in the stratified and unstratified cases are broadly similar.  Stratification somewhat enhances the spreading rate of the passive scalar, and it increases the intensity of the scalar fluctuations in the middle of the layer by about 15\%.  Away from the middle of the layer, though, the scalar is more intermittent in the stratified case.  Nevertheless, the flux profiles scaled by the transverse r.m.s.\ velocity and the peak scalar intensity are very similar once the scalars have had time to come into quasi-equilibrium with the velocity, which takes a somewhat longer time in the stratified case.

In stark contrast, stratification almost stops passive scalar mixing in the vertical.  More precisely, it curtails stirring in the vertical so that the width of the PSML increases until it is proportional to the vertical integral length of the horizontal velocity.  This integral length, in turn, is limited by the constraint that the vertical Froude number will be $O(1)$ \citep{billant01}.  This integral length is almost constant for the duration of the current simulations after about one buoyancy period, but it can be expected to eventually decay in time \citep[c.f.][figure 3]{debk19}.  Since the width of the passive scalar layer cannot decrease in time, this suggests a late-time regime beyond the current scope.

Several methods for modelling the scalar flux are demonstrated.  If the mean scalar profile is known then an eddy-diffusivity model with a single coefficient is effective over the range of times in the simulation.  If the scalar profile is assumed to be an error function, then a two-coefficient model is effective at later times once the scalar has come into quasi-equilibrium with the velocity field.  When this is the case, the length scale of the passive scalar and the length scale of the velocity field are proportional and modelling is straightforward.

The current study is for a simulated flow in which the Prandtl number of the active and passive scalars is 0.7.  Recent studies show the importance of reverse buoyancy flux in stratified flows with $Pr > 1$ for the active scalar \citep{okino19,legaspi20,riley23,bragg24a,bhattacharjee26}.  Since this flux changes the decay rate of the flow relative to that with $Pr=1$, it can be expected to change the mixing rate of the passive scalar.

\vspace*{1em}
This research used resources at the Oak Ridge Leadership Computing Facility at the Oak Ridge National Laboratory, which is supported by the Office of Science of the U.S. Department of Energy under Contract No. DE-AC05-00OR22725.
This work was supported by a grant from the Simons Foundation [SFI-MPS-SRM-00005157, RW].


\bibliography{bib}

\end{document}